\documentclass{cimento}

\usepackage{graphicx}  
\title{Long Gamma Ray Bursts from Quark Stars}
\author{P.~Haensel \atque  J.L.~Zdunik}
\instlist{
\inst{} N. Copernicus Astronomical Center, 00-716 Warszawa, Poland
}

\PACSes{\PACSit{97.60.Jd}{Neutron stars}\PACSit{98.70.Rz}{gamma-ray sources; gamma-ray bursts.}
}

\begin{document}

\maketitle

\begin{abstract}

If strange quark matter (SQM) is the true ground state of
hadronic matter, then conversion of neutron stars (NS) into quark stars (QS)
could release some $10^{53}$\,erg. We describe a scenario of burning
of a NS into a hot, differentially rotating QS. Emission of released non-baryonic
energy through the QS surface is discussed.  The r{\^ol}e  of  magnetobuoyancy
of SQM is mentioned. The outflow of $\gamma e^+e^-$ lasting
for up to $\sim 1000$\,s could be at the origin of long GRBs. Advantages
of hot, differentially rotating QS as an inner engine of long GRBs are reviewed.

\end{abstract}

\section{Introduction}
\label{sect:introd}
Atomic nuclei are droplets of nuclear matter. From the point of view of quark
 structure of matter, nuclear matter is composed of triplets of
 quarks confined to small bubbles of the QCD vacuum. These confined triplets are
 $uud$=protons and  $ddu$=neutrons. Nuclei and electrons form
atoms, and the minimum energy per nucleon  at zero pressure for atomic matter
(thermal contribution being negligible) is reached for a
body-centered cubic crystal of $^{56}{\rm Fe}$,
with energy per nucleon (including nucleon rest energy)
$E(^{56}{\rm Fe})=930.4~$MeV. But is
this a true ground state of matter built of quarks? A
hypothesis that the {\it true} ground state of
{\it hadronic matter}
differs from $^{56}{\rm Fe}$,  and is actually much denser
{\it quark matter},
was advanced since 1970s \cite{Bodmer71,Witten84}(also
Terazawa (1979), as quoted in \cite{Tera89}). In this true ground state,
self-bound at $P=0$, nuclear matter is replaced
by  a plasma  of the $u$, $d$, and $s$ quarks, the presence
of the $s$ quarks being crucial for lowering the energy due to the Pauli
principle. The baryon number of  a drop  of the $uds$ plasma of $N_{\rm q}$
quarks  is $A=N_{\rm q}/3$, and for a
sufficiently large (so that surface and Coulomb energies are much
smaller than the bulk one $\propto A$)
drop of $uds$ plasma
$E(uds)<930.4$\,MeV.
In contrast to atomic nuclei, the size of self-bound droplets of
strange quark matter (SQM)  is not limited by Coulomb forces, and for
$A\sim 10^{57}$  they become huge spheres of SQM called  {\it quark
stars}(QS) \cite{Witten84}. Structure and astrophysics of QS was first
studied in detail  in \cite{HZS86,AFO86}. In contrast to neutron
stars (NS), which are bound by gravitation, QS are bound  by the QCD
forces.  However, for $A\sim 10^{58}$ the quark star mass
$M({\rm QS}) > M_\odot$, while its radius $R\sim 10$\, km. Under such
conditions, gravitation becomes important.  Space-time curvature
implies then the  existence of maximum allowable mass for QSs, which
turns out to be quite similar to that predicted for NSs, $\sim 2\,M_\odot$.

If SQM is indeed the true ground state of hadronic matter, then NS would be metastable
with respect to conversion NS$\longrightarrow$ QS,
the conversion process releasing $\sim 10^{53~}\,$erg. As we will argue,
a newly born QS is a promising source of an ultrarelativistic energy outflow
 which could then  produce   long gamma ray burst (GRB). Since  1986  many papers
 on GRBs from QS appeared and many different scenarios exploiting unique properties
 of QS were proposed (see, e.g.,
 \cite{HPA91,ChengDai96,ChengDai98a,DaiLu98,ChengDai98b,
 BombaciDatta00,WangDai00,OuyedDD02,Lugones2002,
 OuyedSannino02,Berezhiani02,Berezhiani03,DragoLP04a,
 DragoLP04b}).
Here we discuss a specific model of QS as an inner
 engine of long GRBs, stressing its advantages over standard models involving NS.
 Applications to observed long GRBs are presented by Drago,
 Pagliara, and Parenti \cite{DPP_Venice} in these proceedings.
\section{Conversion of neutron star into quark star}
\label{sect:NS-QS}
Conversion NS$\longrightarrow$QS is initiated by nucleation of
a droplet of SQM in nuclear matter near the NS center. This process
is catalyzed by high density and temperature, and could occur near the
center of a newly born NS, formed in gravitational collapse of a core of
a massive star. It could also be triggered in an accreting NS, after its central
density reaches some critical value.

A typical expected latent heat associated with NS$\longrightarrow$QS
 is $Q\sim 50$ MeV/nucleon.
Predicted energy release is therefore $E_{\rm conv}({\rm NS}\longrightarrow {\rm QS})
\sim 10^{53}$\, erg.  This energy release can be re-vitalized by sporadic
accretion.  As the density profile within QS is very different from
that in NS, conversion implies strong differential rotation of a newborn
QS. Typical energy contained in differential rotation (calculated
at constant $A$ and angular momentum $J$)  is  $\Delta E_{\rm \small diff.rot}=
c^2(M_{\rm \small diff.rot}-M_{\rm \small rigid})_{A,J}\sim
10^{52}$\, erg. All in all, the thermal energy,  plus the
 energy in differential rotation of a newly born
QS,  form a huge energy reservoir to generate $\gamma e^+e^-$
fireball with Lorenz factor $\Gamma>100$, which could be at the origin of a long GRB.

Let us consider kinetics of burning of a NS into a QS.
Neutron star is basically composed of a liquid core containing some
99\% of stellar mass and a partially solid crust. For simplicity, we will
consider a core composed of nucleons and  electrons.
 A nucleated droplet of SQM  at the NS center grows by  absorption of
 nucleons, which dissociate into quarks,
$n\longrightarrow u +d +d $,~~$p\longrightarrow d + u +u$.
Then quark matter   equilibrates via  $u+d\longrightarrow s+u$,
~~$u+e\longrightarrow d+\nu_e$. Both processes are highly
exothermic, releasing a total of $\sim 50\,$ MeV/nucleon. The SQM front
progresses outwards. This front can be shown to be  convectively unstable and
consequently progresses as a  {\it strong deflagration},
converting  baryon core into quark core rapidly but still
subsonically, never producing a detonation \cite{DragoPP05}.
The  bottom of the NS crust, containing nuclei, is reached in 10 ms.
The quark core is heated to $\sim 5\times 10^{11}$~K and is  opaque
to neutrinos, which are trapped inside it.

Let us first consider the burning of NS crust into SQM assuming a
diffusive regime. The temperature behind the SQM front is
$T_{\rm \small Q}\sim 10^{11}$~K, and the thickness of the burning layer
where the non-equilibrium reactions take place in quark matter (on the
Q-side), $\delta_{\rm \small Q}$, is much thinner than the
layer of neutron star which is preheated, molten, and
convectively mixed with original crust matter. The thickness of this
 convective layer of the crust ahead of the expanding SQM is
 denoted by  $\delta_{\rm N}$. The temperature of the crust
 before preheating is $T_{\rm crust}\sim 10^{9}-10^{10}\,$K,
 while  $\delta_{\rm \small Q}\ll \delta_{\rm N}\sim
1$\, cm. Let us stress that preheating up to $T_{\rm \small Q}$
leads to dissociation of
nuclei into nucleons which greatly facilitates conversion into SQM.
Generally, preheating and convective mixing strongly accelerates
the burning of the crust into SQM.
 Because of the high density gradient within the crust, its total burning
  moves the SQM front only  30 m outwards. We estimate that this takes 0.01s,
so that the SQM front moves at 3 km/s only.  Because the temperature at the
SQM front is $T_{\rm \small Q}\sim 10^{11}~K$, the very outer layer of
NS,  of mass   $M_{\rm ej}=4\pi R^2 P_\gamma/g$ where
$g\sim 3\times 10^{14}\,{\rm cm/s^2}$,  will be ejected by the photon
pressure. Under prevailing conditions, $M_{\rm ej}\sim 10^{-6}M_\odot$.
\section{Ultrarelativistic energy outflow from the quark star surface}
\label{sect:Eoutflow}
\begin{figure}
\begin{center}
{\includegraphics[width=9cm]{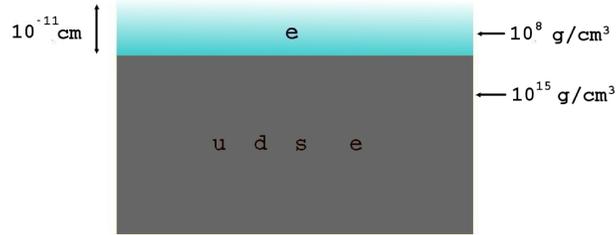}}
\end{center}
\caption{Structure of the quark star surface.}
\label{fig:QSsurface}
\end{figure}
\begin{figure}
\begin{center}
{\includegraphics[width=5.5cm]{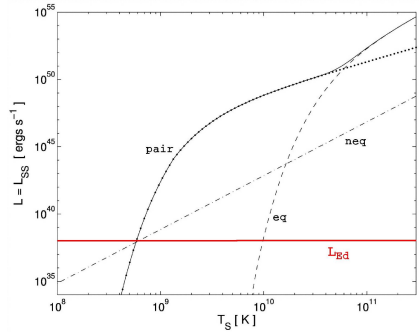}}
\end{center}
\caption{Different contributions to photon luminosity of bare quark star
of $R=10$~km versus surface temperature  $T_{\rm \small S}$.
Spherical symmetry is assumed. For the further
explanation see the text. Based
on \cite{Aksenov2003}}.
\label{fig:Lgamma}
\end{figure}
Quarks, of number density $n_{\rm q}\sim 10^{39}~{\rm cm^{-3}}$,
 are confined  in a huge bubble
 ($R\sim 10$~km) of the QCD vacuum (by strong interaction).
This results in a very sharp SQM  surface, of thickness of the
order of the range of strong interactions $\sim 10^{-13}$\,cm.
However, electrons are not interacting strongly. Their number density in SQM
is $n_e\sim 10^{-4}n_{\rm q}$ \cite{HZS86,AFO86}
and they are bound to quarks (which have
a net positive charge)  by electric forces.
This results in an "electrosphere" of electron gas, of
thickness $10^{-11}$\,cm, extending above the SQM surface
(\cite{AFO86}, Fig.\, \ref{fig:QSsurface}).
The surface layer of a newly born QS is very hot, with
$T_{\rm\small  S}\sim 10^{10}-10^{11}$\,K.
We assume that this does not lead
to a significant evaporation of nucleons from QS. This assumption is valid when
the binding energy of a nucleon in SQM,  $W_N$, satisfies
$W_N\gg k_{\rm B}T_{\rm \small S}$.

 The superdense SQM surface of temperature $T_{\rm\small  S}\sim 10^{10}-10^{11}K$ is
 a very efficient emitter of photons and $e^+e^-$ pairs
 (see \cite{Aksenov2003} and references therein).
 Notice, that as quarks are bound not
 by gravitation but by the (strong) QCD forces,
 and there is no atmosphere, but an electrosphere of a thickness
 of $\sim 10^{-11}$\,cm hold by huge electric forces ($10^{18}$\,V/cm),
 the photon flux emerging
 from QS surface is not  bounded  by the Eddington limit. There are
 three  main mechanisms of photon and $e^+e^-$ pair emission.
 Their contribution to the photon-pair luminosity of a QS
 are:
\par
\parindent 0pt
  ({\bf eq}) Equilibrium transverse plasmons, propagating within $udse$ plasma,
  and emerging as photons outside SQM. As transverse plasmons cannot propagate
  below plasma frequency  $\omega_{\rm plas}\sim 20~{\rm MeV}/\hbar$, this component of the
 photon flux, denoted by {\tt eq} in Fig.\ \ref{fig:Lgamma}, is
 completely negligible for $T_{\rm \small S}<10^{10}$\,K.\par
({\bf pair}) Pair emission and
 annihilation. Pairs are efficiently formed in the huge electric field in the surface
 layer, and for  $5\times 10^8~{\rm K}<T_{\rm \small S}<5\times 10^{10}~{\rm K}$
 give a dominant contribution to the energy outflow, see Fig.\
 \ref{fig:Lgamma}.\par
 ({\bf neq}) Photons produced in Bremsstrahlung  $qq$ and $ee$
 processes in the surface layer.\par
 \parindent 21pt
Finally, we have  $\nu\overline{\nu}\longrightarrow e^+e^-$ above the SQM surface.
Actually, sharp QS surface creates ideal conditions for
efficient  $\nu\overline{\nu}\longrightarrow e^+e^-$ pair production. However, its
contribution, so important for NS, is not important for hot
QS, because even at  $10^{11}\,$K it yields  only $\sim 1\%$ of
the total luminosity.

\section{Differential rotation and magnetobuoyancy}
\begin{figure}
\begin{center}
\includegraphics[height=4.5cm]{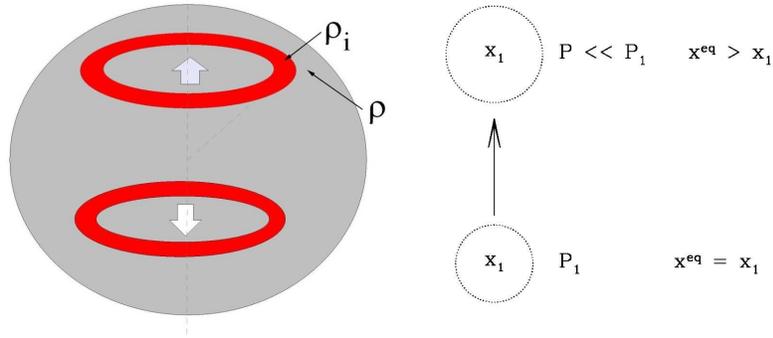}
\end{center}
\caption{{\it Left panel.}
Floating of magnetized toroids due to their buoyancy. Quark star
rotates around a vertical axis. $\rho_i$ and $\rho$
are matter density in the toroid
and the ambient matter density, respectively.
{\it Right panel.} Figure explaining the antibuoyancy due to stratification.
}
\label{fig:Btor-abuo}
\end{figure}

QS has a nearly constant density, while in NS  we have
$\rho_{\rm center}/\rho_{\rm surf}\sim
10^{14}$. Therefore, a rigidly rotating NS converts into a strongly
differentially rotating hot QS.  As the electrical conductivity of SQM is
huge, this makes ideal conditions to generate toroidal magnetic field
$B_{\rm tor}$ by winding an initial poloidal $B_{\rm pol}$,
amplified later  by the magneto-rotational instability. A
schematic picture of magnetized toroids inside a rotating QS
is shown in Fig.\ \ref{fig:Btor-abuo}.
Pressure equilibrium requires $P=P_{\rm i}+{B^2/(8\pi)}$
which implies matter density difference
$\rho_{\rm i}-\rho=-{B^2/(8\pi c_{\rm s}^2)}~$,
where $c_{\rm s}\approx 0.6\, c$ is the sound speed in quark
matter.
Effective total mass-energy density deficit within the toroid,
relative to the ambient medium,  is
$
(\Delta\rho)_{\rm b}=-\left(c^2-c_{\rm s}^2\right)
{B^2/(8\pi c_{\rm s}^2)}~.
$
A magnetized toroid, pushed up by buoyancy,  floats along the
rotation axis towards  the QS surface, Fig.\ \ref{fig:Btor-abuo}.
\subsection{Stratification and anti-buoyancy: electrons
present}
Electron fraction in SQM $x=n_e/n_{\rm b}$ increases outwards
\cite{HZS86,AFO86}. A floating ``magnetized ring'' gets off beta equilibrium.
 At  ambient $P$, a non-equilibrated
quark matter weights more than the
equilibrated one. Therefore,  stratification opposes
buoyancy of a magnetized SQM (Fig.\ \ref{fig:Btor-abuo}).
Consider an element of SQM initially at pressure $P_1$ and with
equilibrium composition $x_1=x^{\rm eq}(P_1)$, moving
upwards, with fixed $x=x_1$,  to $P\ll P_1$ through the
equilibrated medium with $x=x^{\rm eq}(P)$ (Fig.\
\ref{fig:Btor-abuo}). The anti-buoyancy factor is then defined by
$f_{\rm ab}(P_1;P)=[{\rho(P,x_1)-\rho^{\rm eq}(P)]/\rho^{\rm eq}(P)}$~,
where $\rho^{\rm eq}(P)\equiv \rho(P,x^{\rm eq}(P))$. By
construction, $f_{\rm ab}(P_1;P_1)=0$
\subsection{Antibuoyancy for quark stars and neutron stars}
Factor $f_{\rm ab}^{\rm QS}$ is very sensitive to the
mass of strange quark $m_s$. At $T=0$ we get
$f_{\rm ab}=1\times 10^{-7}
\times [{m_s c^2/(100\,{\rm MeV})}]^{7.3}$ \cite{HZ06}.
 Thermal effects increase $f_{\rm ab}$. Let us define $T_{11}=T/10^{11}~K$.
For $T_{11}>1$ and at fixed $m_s$, thermal effects dominate,
and one gets $f_{\rm ab}^{\rm QS}\propto
T_{11}^2$ \cite{Drago06}. The values for NS are many orders of
magnitude  larger, $f_{\rm ab}^{\rm NS}\sim 10^{-2}$
\cite{KR98,HZ06}.
\subsection{Maximum $B=B_{\rm f}$ halted by stratification}
Floating of magnetized toroid of SQM is halted
 if $\left(\Delta\rho\right)_{\rm ab}=
 \rho f_{\rm ab}>\left(\Delta\rho\right)_{\rm b}$.
The toroid will float towards the QS surface, along the rotation axis,  if
$B>B_{\rm f}=\sqrt{8\pi f_{\rm ab}\rho c_{\rm s}^2}$
(Fig.\ \ref{fig:Btor-abuo}). We have
$B_{\rm f}\approx 10^{15}~(f_{\rm ab}/10^{-6})^{1/2}$\,G.
Therefore, for QS with $T_{11}>1$ and  $m_s=200~{\rm MeV}/c^2$ we
obtain  $B_{\rm  f}^{\rm QS}=5\times 10^{15}T_{11}~G$ \cite{HZ06}.
For NS the critical value is  $B_{\rm  f}^{\rm NS}=10^{17}~G$
\cite{KR98}.\par
After $B^{\rm QS}_{\rm f}$ is reached, the magnetized toroid
floats to the QS surface. The  ultrarelativistic  $\gamma e^+e^-$ outflow from the
QS surface follows. Using the arguments
of \cite{KR98}, applied originally to NSs, one can show that the duration of the
outflow, resulting from many cycles of the toroid generation
and floating, can be as long as 1000 s.

In the case when QS is on the CFL superducting state, there is
a possibility of an CFL-enforced absence of electrons and no
stratification. Then buoyancy is not opposed, and
the floating of B-toroid and winding up are simultaneous.
However, the floating is so slow, that the whole process of
transport of frozen-in magnetic field to the QS surface, until
$E^{\rm diff}_{\rm rot}\sim 10^{52}$\,~ erg is exhausted, takes
hundreds of seconds.
\section{Quark star inner engine vs.  baryonic one}
Let us summarize specific features of the inner engine of long
GRBs based on the NS$\longrightarrow$QS transition. This
transition can take place in a newly born NS after a SNIc
explosion, or in a spinning down or accreting NS. In the latter case,
there is no SN accompanying the long GRB.

The process starts with a rapid conversion NS$\longrightarrow$ QS which
can produce $E_{\rm therm} \sim 10^{53}$\,erg  and
 $E^{\rm diff}_{\rm rot}\sim 10^{52}\,$erg on a timescale $\sim 0.01~
 s$. There is  no detonation, and matter ejection
 $~10^{-6}~M_\odot$ is  insignificant. The quark surface
 acts as a membrane - only non-baryonic
energy can flow through it. There is therefore  no  problem with
Eddington limit for photons, so severe for NSs, where
baryon loading cannot be avoided if $L>L_{\rm Edd}$. Moreover,
the  quark surface is sharp, which makes
$\nu\overline{\nu}\longrightarrow e^-e^+$ process very
efficient (in contrast to NS where this process causes a
strong baryonic wind). A toroidal magnetic field, generated in a
differential rotation, can power a long-time tail of energy
outflow from the QS surface, without baryon pollution.
For NS with $L>>L_{\rm Edd}$ and neutrinosphere deep in stellar
interior, a strong baryon pollution cannot be avoided.

Concluding, a hot differentially rotating QS  could be an efficient
inner engine of long GRBs  (for application of this model to
GRBs see \cite{DPP_Venice} in these proceedings).
 Alas, we do not know whether QS exist.
\acknowledgments We express our gratitude to Bohdan Paczy{\'n}ski,
who suggested  some of the studies described in the present paper.
One of the authors (PH) is deeply grateful to Professor Paczy{\'n}ski
for introducing him into  the gamma-ray bursts astrophysics.
 We are grateful to Alessandro Drago
for collaboration on the conversion of neutron stars into quark stars,
and for reading the manuscript.  This work was supported by the MNiI grant
no. 1P03D-008-27.

\end{document}